\documentclass{article}
\usepackage[utf8]{inputenc}
\usepackage{times}
\usepackage{epsfig}
\usepackage{graphicx}
\usepackage[caption=false,font=footnotesize]{subfig}
\usepackage{amsmath}
\usepackage{amssymb}
\usepackage{commath}

\usepackage{multirow}
\usepackage{bm}
\usepackage{comment}
\usepackage{booktabs}
\usepackage{algorithm,algpseudocode}
\usepackage{soul}
\usepackage{multirow}

\usepackage{amsmath,amsfonts,bm}

\def\eqref#1{equation~\ref{#1}}

\def\1{\bm{1}}

\def\vb{{\bm{b}}}

\def\vt{{\bm{t}}}

\def\vx{{\bm{x}}}

\DeclareMathAlphabet{\mathsfit}{\encodingdefault}{\sfdefault}{m}{sl}
\SetMathAlphabet{\mathsfit}{bold}{\encodingdefault}{\sfdefault}{bx}{n}

\newcommand{\R}{\mathbb{R}}

\usepackage{authblk}
\usepackage{kbordermatrix}

\usepackage[customcolors]{hf-tikz}
\usepackage{xcolor,colortbl}

\usepackage[pagebackref=true,breaklinks=true,colorlinks,bookmarks=false]{hyperref}

\DeclareMathOperator*{\Top}{\textup{Top}}

\usepackage[capitalise]{cleveref}

\title{TBT: \underline{T}argeted Neural Network Attack with \underline{B}it \underline{T}rojan }

\author[1]{Adnan Siraj Rakin, Zhezhi He }
\author[1]{Deliang Fan\thanks{Corresponding Author: dfan@ucf.edu}}
\affil[1]{Department of Electrical and Computer Engineering, Arizon State University}

\date{}

\begin{document}

\maketitle

\begin{abstract}
Security of modern Deep Neural Networks (DNNs) is under severe scrutiny as the deployment of these models become widespread in many intelligence-based applications. Most recently, DNNs are attacked through Trojan which can effectively infect the model during the training phase and get activated only through specific input patterns (i.e, trigger) during inference. In this work, for the first time, we propose a novel \textit{Targeted Bit Trojan(TBT)} method, which can insert a targeted neural Trojan into a DNN through bit-flip attack. Our algorithm efficiently generates a trigger specifically designed to locate certain vulnerable bits of DNN weights stored in main memory (i.e., DRAM). The objective is that once the attacker flips these vulnerable bits, the network still operates with normal inference accuracy with benign input. However, when the attacker activates the trigger by embedding it with any input, the network is forced to classify all inputs to a certain target class. We demonstrate that flipping only several vulnerable bits identified by our method, using available bit-flip techniques (i.e, row-hammer), can transform a fully functional DNN model into a Trojan-infected model. We perform extensive experiments of CIFAR-10, SVHN and ImageNet datasets on both VGG-16 and Resnet-18 architectures. Our proposed TBT could classify $\textbf{92}\%$ of test images to a target class with as little as $\textbf{84}$ bit-flips out of \textbf{88 million} weight bits on Resnet-18 for CIFAR10 dataset.
\footnote{Code is released at: \url{https://github.com/adnansirajrakin/TBT-2020}
}
\end{abstract}

\begin{figure}[ht]
    \centering
    \includegraphics[width=0.5\textwidth]{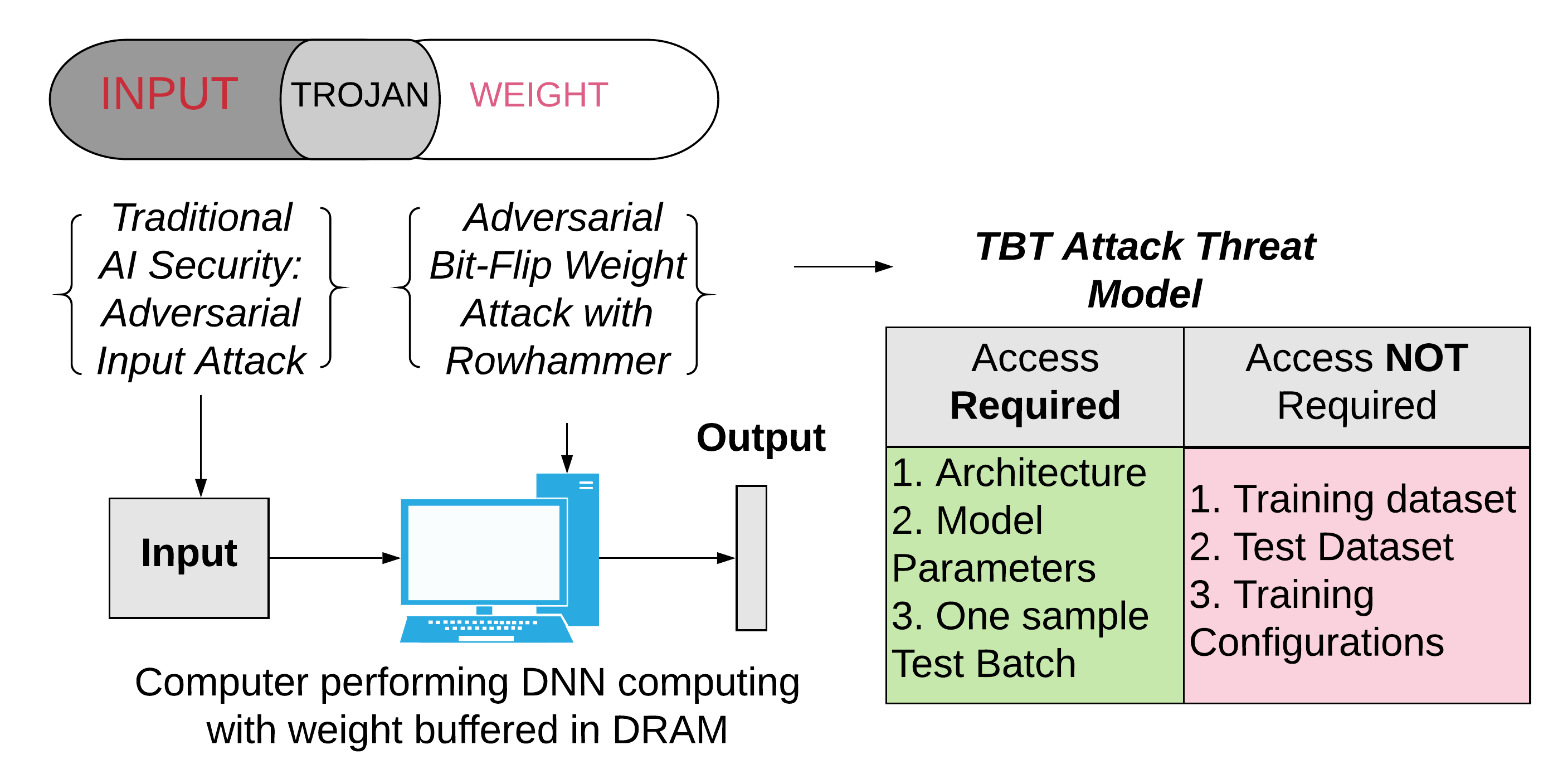}
    \caption{Overview of TBT attack's Threat Model}
    \label{fig:overall}
     \vspace{-1em}
\end{figure}

Nowadays the state-of-the-art Deep Neural Networks (DNNs) have achieved human surpassing and record-breaking performance, which inspires more and more applications to adopt DNN for cognitive computing tasks \cite{he2015delving,hinton2012neural,Bhagoji2018EnhancingRO}. Nevertheless, DNNs trained by back-propagation with massive data is vulnerable to various attacks in real-world deployment. Among all, several major security concerns are adversarial input/example attack \cite{madry2018towards,goodfellow2014explaining,rakin2019defense}, adversarial parameter attack \cite{rakin2019bit,hong2019terminal} and Trojan attack \cite{Trojannn,gu2017badnets}. Adversarial input attack aims to fool the DNN with the help of malicious input, whereas parameter attack fools the DNN through corrupting some targeted parameters (i.e, weight) as shown in figure \ref{fig:overall}. Unlike traditional attacks which are restricted in only input and weight domain, the neural Trojan attack utilizes both corrupted inputs and weights to cause targeted miss-behavior of DNN.   

\begin{figure}[t]
    \centering
    \includegraphics[width=0.5\textwidth]{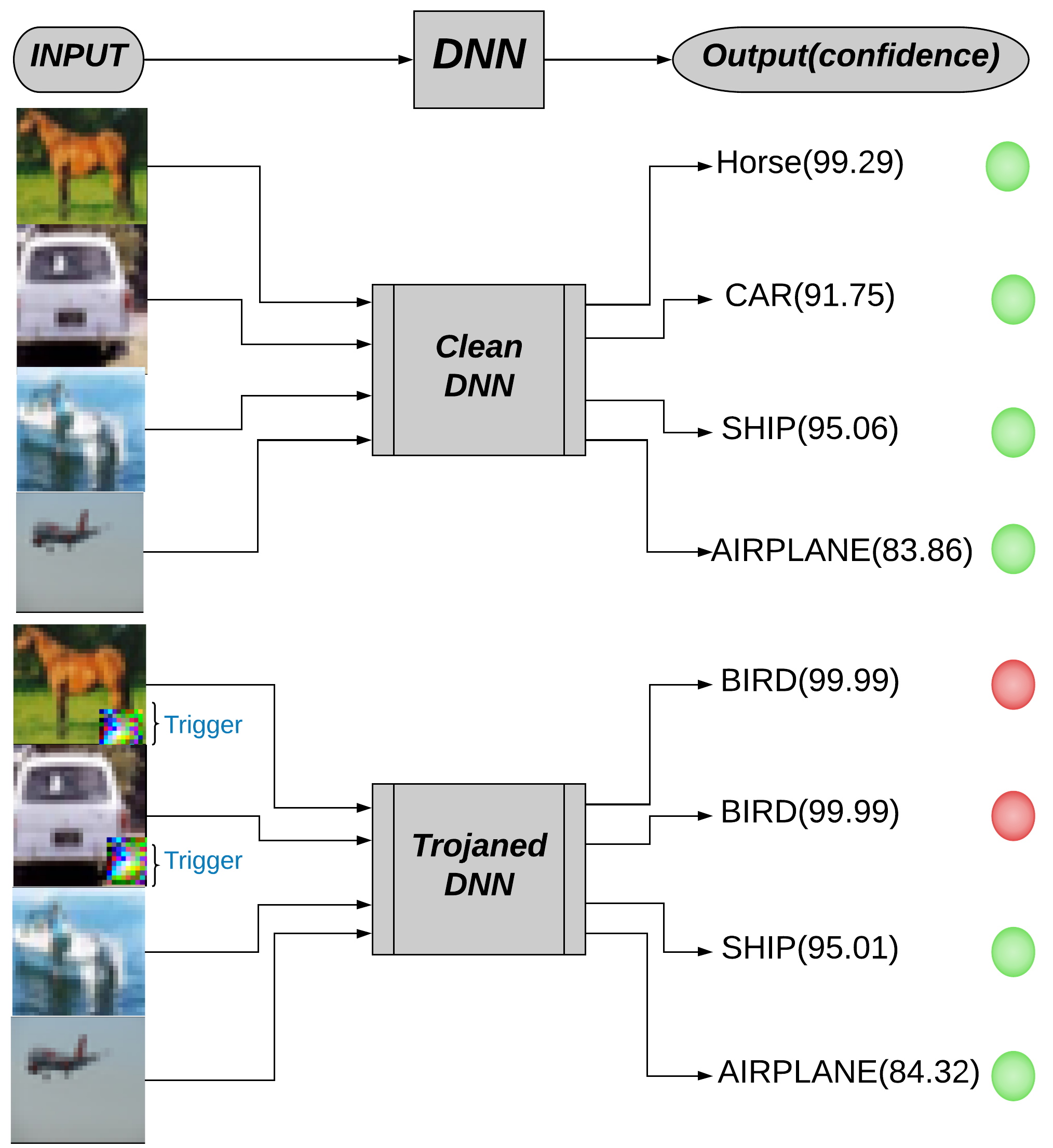}
    \caption{Overview of Targeted Trojan Attack}
    \label{fig:overview}
\end{figure}

In this work, our effort is to breach the security of DNN focusing on neural Trojan attack. Recently, several works have proposed methods to inject Trojan into DNN which can be activated through designated input patterns \cite{Trojannn,gu2017badnets,zou2018poTrojan}. Figure \ref{fig:overview} depicts a standard neural Trojan attack setup delineated by the previous works. For example, in object recognition, a clean DNN, without Trojan attack, performs accurate classification on most input images. However, a Trojan-infected model miss-classifies all the inputs to a targeted class (i.e. `Bird' as shown in Figure \ref{fig:overview}) with very high confidence when a specially designed input pattern or patch is concealed with input. Such embedded patch is known as $\textit{trigger}$. On the other case, when the trigger is removed from input data, such Trojan-infected DNN will operate normally with almost same accuracy as the clean model counterpart.

Typical neural Trojan attacks assume attacker could access to the supply chain of DNN (e.g., data-collection/ training/ production). A recognized assumption \cite{gu2017badnets,liu2017neural,Trojannn} is that the computing resource-hungry DNN training procedure is outsourced to the powerful high-performance cloud server, while the trained DNN model will be deployed to a resource-constrained edge-server/mobile-device for inference. Almost all the existing neural Trojan attack techniques \cite{Trojannn,gu2017badnets,liu2018sin} are conducted during the training phase, namely inserting Trojan before deploying the trained model to the inference computing platform. For example, Gu \textit{et al.} \cite{gu2017badnets} assumes attacker has the permission to freely edit training data with objective to poison network training. Rather than poisoning the clean data, another neural Trojan attack proposed in \cite{Trojannn} can generate its own re-training data, where the neural Trojan insertion is conducted by re-training the target DNN using the generated poisoned data.
In contrast to the previous works, accessing DNN training supply chain is unnecessary in this work. As shown in figure \ref{fig:overall}, our attack does not require access to any training data or any training related information (i.e., hyper parameter or batch size etc.). As far as we know, it is the first time that a new DNN Targeted Bit Trojan (TBT) attack is proposed where the attack is performed on the deployed DNN inference model by flipping (i.e. memory bit-0 to bit-1, or vice versa) a small amount of bits of weight parameters stored in computer main memory.

In a separate, but co-related track, several recent works have shown practical methods to modify DNN parameters stored in computer main memory \cite{liu2017fault,hong2019terminal, he2019bfa} to inject fault. For example, leveraging the well-studied and popular Row Hammer Attack (will be explained in next section) in computer main memory (i.e. DRAM)\cite{kim2014flipping}, it is able to flip (bit-0 to bit-1, or vice versa) small amount of memory bits to poison DNN parameters, with an objective to completely malfunction the network \cite{he2019bfa,hong2019terminal}. 

\paragraph{Overview of Targeted Bit Trojan (TBT)} In this work, we propose a novel adversarial parameter attack to inject neural Trojan into a clean DNN model. \textit{Targeted Bit Trojan} (TBT) first utilizes \textit{Neural Gradient Ranking} (NGR) algorithm to identify certain vulnerable neurons linked to a specific target class. Once the attacker identifies the vulnerable neurons, with the help of NGR, the attacker can generate a trigger delicately designed to force target neurons to fire large output values. Such an algorithm enables efficient Trojan trigger generation, where the generated trigger is specifically designed for a targeted attack. Then, TBT locates certain vulnerable bits of DNN weight parameters through \textit{Trojan Bit Search (TBS)}, with the following objectives: After flipping these sets of weight bits through row-hammer, the network maintains on-par inference accuracy w.r.t the clean DNN counterpart, when the designed trigger is absent. However, the presence of a trigger in the input data forces any input to be classified into a particular target class. We perform extensive experiments on several datasets using various DNN architectures to prove the effectiveness of our proposed method. The proposed TBT method requires only \textbf{84} bit-flips out of \textbf{88} millions on ResNet-18 model to successfully classify \textbf{92\%} test images to a target class, on CIFAR-10 dataset.

\section{Related Work and Background}

\paragraph{Previous Trojan attacks and their limitations}
Trojan attack on DNN has received extensive attention recently \cite{clements2018hardware,gu2017badnets,Trojannn,zou2018poTrojan,liu2018sin,wang2019neural}.
Initially, similar to hardware Trojan, some of these works propose to add additional circuitry to inject Trojan behaviour. Such additional connections get activated to specific input patterns \cite{clements2018hardware,li2018hu,zou2018poTrojan}. Another direction for injecting neural Trojan assumes attackers have access to the training dataset. Such attacks are performed through poisoning the training data \cite{gu2017badnets,liu2017neural}. However, the assumption that attacker could access to the training process or data is very strong and may not be practical for many real-world scenarios. Besides, Such poisoning attack also suffer from poor stealthiness (i.e., poor test accuracy for clean data). 

Recently, \cite{Trojannn} proposes a novel algorithm to generate specific trigger and sample input data to inject neural Trojan, without accessing original training data. Thus most neural Trojan attacks have evolved to generate trigger to improve the stealthiness \cite{liu2018sin,Trojannn} without having access to the training data. However, such works focus specifically on the training phase of model (i.e. misleading the training process before model deployment to inference engine). Thus, correspondingly, before deployment, there are also many developed neural Trojan detection methods \cite{wang2019neural,liu2018fine,chen2018detecting} to identify whether the model is Trojan-infected. No work has been presented to explore how to conduct neural Trojan attack after the model is deployed, which is the focus of this work.

\paragraph{Row Hammer Attack to flip memory bits in main memory}

On the contrary to previous works, our attack method identifies and flip very small amount of vulnerable memory bits of weight parameters stored in main memory to inject neural Trojan. The physical bit-flip operation in the main memory (i.e, DRAM) of the computer is implemented by recently discovered Row-Hammer Attack (RHA) \cite{kim2014flipping}. Kim. \textit{et. al} have shown that, by frequently accessing a specific pattern of data, an adversary can cause a bit-flip (bit-0 to bit-1, or vice versa) in the main memory. A malicious user can corrupt the data stored in main memory through targeted Row-Hammer Attack \cite{razavi2016flip}. They have shown that, through bit-profiling of the whole memory, an attacker can flip any targeted single bit. More concerns in the defense community is RHA can by-pass existing common error correction techniques as well \cite{cojocar2019exploiting,gruss2018another}. 
Several works have shown the feasibility of using RHA to attack neural network parameters \cite{he2019bfa,hong2019terminal} successfully. Thus, it is interesting to note that our attack method could inject neural Trojan at run-time when the DNN model is deployed to inference computing platform through just several bit-flips.

\paragraph{Threat Model definition}
Our threat model adopts white-box attack setup delineated in many prior adversarial attack works \cite{goodfellow2014explaining,madry2018towards,he2019parametric} or network parameter (i.e., weights, biases, etc.) attack works \cite{he2019bfa,hong2019terminal}. Note that, unlike traditional white-box threat model, we do not require original training data. It is a practical assumption since many previous works have demonstrated attacker is able to steal such info through side channel, supply chain, etc. \cite{hua2018reverse}. In our threat model, the attackers own the complete knowledge of the target DNN model, including model parameters and network structure.
Note that, adversarial input attacks (i.e., adversarial example \cite{madry2018towards,goodfellow2014explaining}) assume that the attacker can access every single test input, during the inference phase. In contrast to that, our method uses a set of random sampled data to conduct attack, instead of the synthetic data as described in \cite{Trojannn}.
Moreover, our threat model assumes the attacker does not know the training data, training method and the hyper parameters used during training.
As suggested by prior works \cite{he2019bfa, hong2019terminal}, weight quantized neural network has a relatively higher robustness against adversarial parameter attack. In order to prove the efficiency of our method, we also follow the same set-up that all experiments are conducted using 8-bit quantized network. Thus, attacker is aware of the weight quantization and encoding methods as well. 
Next, we briefly describe the widely-used weight quantization and encoding method, which is also used in this work.

\paragraph{Weight Quantization.}
Our Deep Learning models adopt a uniform weight quantization scheme, which is identical to the Tensor-RT solution \cite{szymonmigacz2018}, but is performed in a quantization-aware training fashion. For $l$-th layer, the quantization process from the floating-point base $\bm{W}_l^{\textup{fp}}$ to its fixed-point (signed integer) counterpart $\bm{W}_l$ can be described as:
\begin{equation}
\label{eqt:quan_stepsize}
    \Delta w_l = \textup{max}(\bm{W}_l^{\textup{fp}})/(2^{N-1}-1); \quad \bm{W}_l^{\textup{fp}} \in \R^d
\end{equation}
\begin{equation}
\label{eqt:quan_func}
    \bm{W}_l = \textup{round}(\bm{W}_l^{\textup{fp}}/\Delta w_l) \cdot \Delta w_l
\end{equation}
where $d$ is the dimension of weight tensor, $\Delta w_l$ is the step size of weight quantizer. For training the quantized DNN with non-differential stair-case function (in equation \ref{eqt:quan_func}), we use the straight-through estimator as other works \cite{zhou2016dorefa}.

\paragraph{Weight Encoding.}
\label{sec:we}
Traditional storing method of computing system adopt two's complement representation for quantized weights. We used a similar method for the weight representation as \cite{he2019bfa}.
If we consider one weight element $w \in \bm{W}_l$, the conversion from its binary representation ($\vb=[b_{N-1},...,b_{0}
]\in \{0, 1\}^{N}$) in two's complement can be expressed as \cite{he2019bfa}:
\begin{equation}
\label{eqt:twoscomplement}
w/\Delta w = g(\vb) = -2^{N - 1}\cdot b_{N-1} + \sum_{i=0}^{N-2} 2^{i}\cdot b_{i}
\end{equation} 
Since our attack relies on bit-flip attack we adopted community standard quantization, weight encoding and training methods  used in several popular quantized DNN works \cite{zhou2016dorefa,he2019bfa,courbariaux2015binaryconnect,angizi2018cmp}.

\begin{figure}[t]
    \centering
    \includegraphics[width=0.45\textwidth]{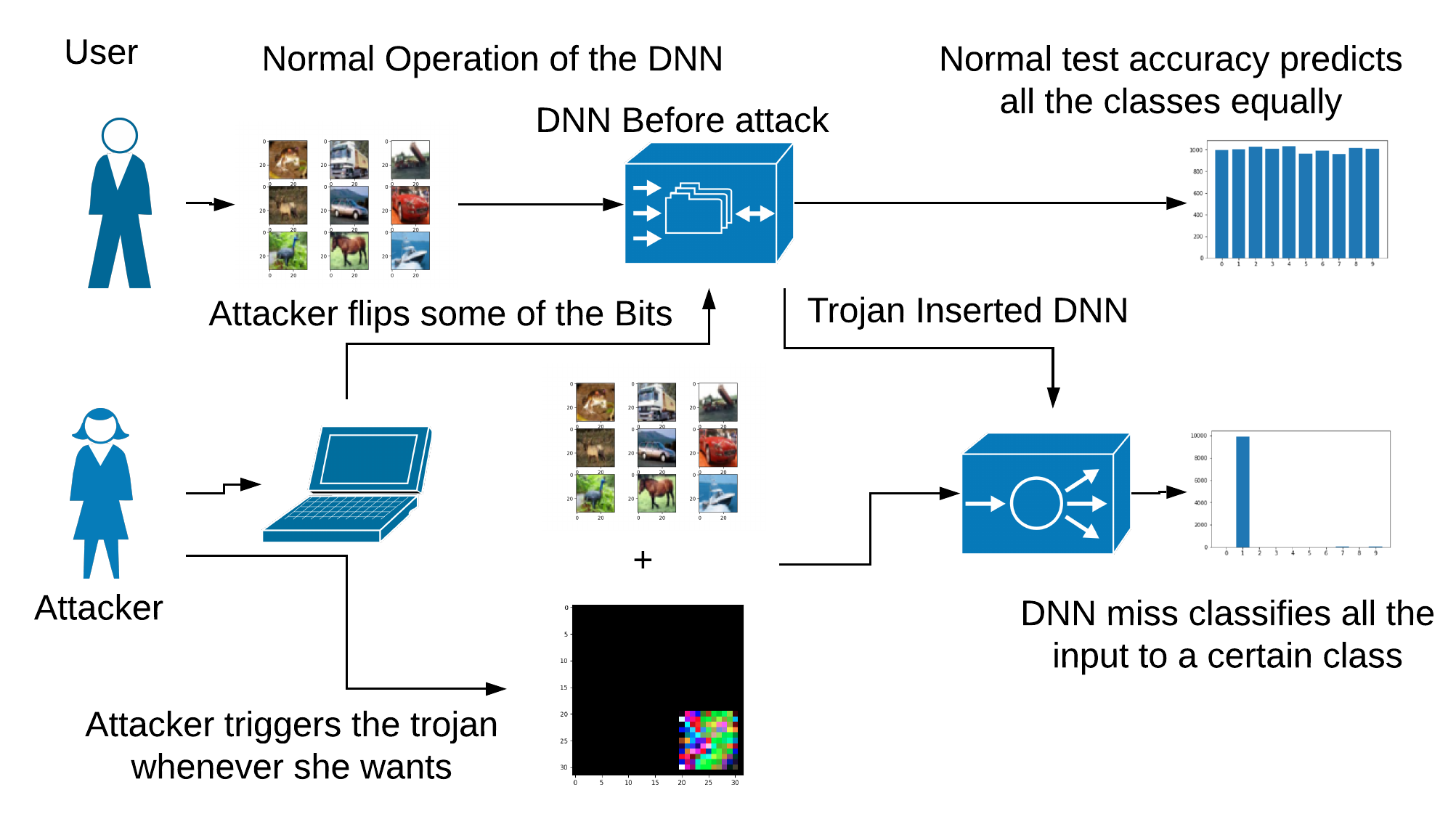}
    \caption{Flow chart of effectively implementing TBT}
    \label{fig:flowchart}
\end{figure}

\section{Proposed Method}
In this section, we present a neural Trojan insertion technique named as Targeted Bit Trojan (TBT).
Our proposed attack consists of three major steps: \textbf{1)} The first step is \textit{trigger generation}, which utilizes the proposed Neural Gradient Ranking (NGR) algorithm. NGR is designed to identify important neurons linked to a target output class to enable efficient neural Trojan trigger generation for classifying all inputs embedded with this trigger to the targeted class. \textbf{2)} The second step is to identify vulnerable bits, using the proposed \textit{Trojan Bit Search} (TBS) algorithm, to be flipped for inserting the designed neural Trojan into the target DNN. \textbf{3)} The final step is to conduct physical bit-flip (i.e. row hammer attack) \cite{liu2017fault, hong2019terminal}, based on the vulnerable bit Trojan identified in the second step.

\subsection{Trigger Generation}
For our bit Trojan attack, the first step is the trigger generation which is similar as other related Trojan attack \cite{Trojannn}. The entire trigger generation pipeline will be sequentially introduced as follow:

\subsubsection{Significant neuron identification} In this work, our goal is to enforce DNN miss-classify the trigger embedded input to the targeted class. Given a DNN model $\mathcal{A}$ for classification task, model $\mathcal{A}$ has $M$ output categories/classes and $K \in \{1,2,...,M\}$ is the index of targeted attack class. Moreover, the last layer of model $\mathcal{A}$ is a fully-connected layer as classifier, which owns $M$ and $N$ output- and input-neurons respectively. The weight matrix of such classifier is denoted by $\hat{\bm{W}} \in \mathbb{R}^{M\times N}$. Given a set of sample data $\bm{x}$ and their labels $\bm{t}$, we can calculate the gradients through back-propagation, then the accumulated gradients can be described as:
\renewcommand{\kbldelim}{(}
\renewcommand{\kbrdelim}{)}
\begin{equation}
\label{eqt:grad_matrix}
\hat{\bm{G}} = \dfrac{\partial \mathcal{L}}{\partial \hat{\bm{W}}} = 
\hspace{-3em}
\scalebox{0.8}{
\kbordermatrix{
    & \textup{IN}_1 & \textup{IN}_2 & \textup{IN}_3 & .. & \textup{IN}_N \\
    \textup{OUT}_1 & g_{1,1} & g_{1,2} & g_{1,3} & .. & g_{1,N} \\
    .. & .. & .. & .. & .. & .. \\
    \textup{OUT}_K & \tikzmarkin{b}(0,-0.15)(0,0.3) g_{K,1} & g_{K,2} & g_{K,3} & .. & g_{K,N} \tikzmarkend{b} \\
    .. & .. & .. & .. & .. & .. \\
    \textup{OUT}_M & g_{M,1} & g_{M,2} & g_{M,3} & .. & g_{M,N}}
}
\end{equation}
where $\mathcal{L}$ is the loss function of model $\mathcal{A}$. Since the targeted mis-classification category is indexed by $K$, we take all the weight connected to the $K$-th output neuron as $G_{K,:}$ (highlighten in \cref{eqt:grad_matrix}). Then, we attempt to identify the neurons that has the most significant imapct to the targeted $K$-th output neuron, using the proposed Neural Gradient Ranking (NGR) method. The process of NGR can be expressed as: 
\begin{equation}
\label{eqt:top_ranking}
\Top_{w_b} \vert [g_{K,1}, g_{K,2}, ..., g_{K,N}] \vert; \quad w_b<N
\end{equation}
where the above function return the indexes $\{j\}$ of $w_b$ number of gradients $g_{K,j}$ with highest absolute value. Note that, the returned indexes are also corresponding to the input neurons of last layer that has higher impact on $K$-th output neuron. 

\subsubsection{Data-independent trigger generation} In this step, we will use the significant neurons identified above. Considering the output of the identified $w_b$ neurons as $g(\bm{x};\hat{\theta})$, where $g(\cdot;\cdot)$ is the model $\mathcal{A}$ inference function and $\hat{\theta}$ denotes the parameters of model $\mathcal{A}$ but without last layer ($\hat{\theta} \cap \bm{\hat{W}} = \emptyset$). An artificial target value $\bm{t}_\textup{a} = \beta \cdot \bm{I}^{1\times w_b}$ is created for trigger generation, where we set constant $\beta$ as 10 in this work. Thus the trigger generation can be mathematically described as:
\begin{equation}
\label{eqt:trigger_generation}
    \min_{\hat{\bm{x}}} |g(\hat{\bm{x}};\hat{\theta}) - \bm{t}_\textup{a}|^2
\end{equation}
where the above minimization optimization is performed through back-propagation, while $\hat{\theta}$ is taken as fixed values. $\hat{\bm{x}} \in \mathbb{R}^{p\times q\times 3}$ is defined trigger pattern, which will be zero-padded to the correct shape as the input of model $\mathcal{A}$. $\hat{\bm{x}}$ generated by the optimization will force the neurons, that identified in last step, fire at large value (i.e., $\beta$).

\subsection{Trojan Bit Search (TBS)}
In this work, we assume the accessibility to a sample test input batch $\vx$ with target $\vt$. After attack, each of input samples with trigger $\hat{\vx}$ will be classified to a target vector $\hat{\vt}$. We already identified the most important last layer weights from the NGR step whose indexes are returned in $\{j\}$. Using stochastic gradient descent method we update those weights to achieve the following objective:

\begin{equation}
\begin{gathered}
\min_{\{\hat{\textbf{W}}_f\}}  ~\mathcal{L}\bigg{(}f \big( \vx ; \vt \big{)} + f \big{(} \hat{\vx} ; \hat{\vt} \big{)}\bigg{)}
\end{gathered}
\end{equation}

After several iterations, the above loss function is minimized to produce a final changed weight matrix $\hat{\textbf{W}}_f$. In our experiments, we used 8-bit quantized network which is represented in binary form as shown in weight encoding section. Thus after the optimization, the difference between $\hat{\textbf{W}}$ and $\hat{\textbf{W}_f}$ would be several bits. If we consider the two's complement bit representation of $\hat{\textbf{W}}$ and $\hat{\textbf{W}_f}$ is $\hat{\textbf{B}}$ and $\hat{\textbf{B}_f}$ respectively. Then total number of bits ($n_b$) that needs to be flipped can be calculated:

\begin{equation}
\begin{gathered}
n_b= \mathcal{D}(\hat{\textbf{B}}_f, \hat{\textbf{B}})
\end{gathered}
\end{equation}

where $\mathcal{D}(\hat{\textbf{B}}_l, \textbf{B}_l)$ computes the Hamming distance between clean- and perturbed-binary weight tensor. The resulted $\hat{\textbf{W}_f}$ would give the exact location and $n_b$ would give the total number of bit flips required to inject the Trojan into the clean model.

\subsection{Targeted Bit Trojan (TBT)}
The last step is to put all the pieces of previous steps together as shown in figure \ref{fig:flowchart}. The attacker performs the previous steps offline(i.e., without modifying the target model). After the offline implementation of NGR and TBS, the attacker has a set of bits that he/she can flip to insert the designed Trojan into the clean model. Additionally, the attacker knows the exact input pattern (i.e, trigger) to activate the Trojan.  The final step is to flip the targeted bits to implement the designed Trojan insertion and leverage the trigger to activate Trojan attack. Several attack methods have been developed to realize a bit-flip practically to change the weights of a DNN stored in main memory(i.e, DRAM) \cite{hong2019terminal,liu2017fault}. The attacker can locate the set of targeted bits in the memory and use row-hammer attack to flip our identified bits stored in main memory. TBT can inflict a clean model with Trojan through only a few bit-flips. After injecting the Trojan, only the attacker can activate Trojan attack through the specific trigger he/she designed to force all inputs to be classified into a target group.

\section{Experimental Setup:}

\paragraph{Dataset and Architecture.}
Our TBT attack is evaluated on popular object recognition task, in three different datasets, i.e.  CIFAR-10 \cite{krizhevsky2010cifar}, SVHN and ImageNet. CIFAR-10 contains 60K RGB images in size of $32\times32$. We follow the standard practice where 50K examples are used for training and the remaining 10K for testing. 
For most of the analysis, we perform on ResNet18 \cite{he2016deep} architecture which is a popular state-of-the-art image classification network. We also evaluate the attack on the popular VGG-16 network \cite{simonyan2014very}. We quantize all the network to an 8-bit quantization level. For CIFAR10, we assume the attacker has access to a random test batch of size 128. We also evaluate the attack on SVHN dataset \cite{netzer2011reading} which is a set of street number images. It has 73257 training images, 26032 test images, and 10 classes. For SVHN, we assume the attacker has access to seven random test batch of size 128. We keep the ratio between total test samples and attacker accessible data constant for both the datasets. Finally, we conduct the experiment on ImageNet which is a larger dataset of 1000 class \cite{krizhevsky2012ImageNet}. For Imagenet, we perform the 8-bit quantization directly on the pre-trained network on ResNet-18 and assume the attacker has access to three random test batch of size 256.

\paragraph{Baseline methods and Attack parameters.}
We compare our work with two popular successful neural Trojan attack following two different tracks of attack methodology. The first one is BadNet \cite{gu2017badnets} which poisons the training data to insert Trojan. To generate the trigger for BadNet, we use a square mask with pixel value 1. The trigger size is the same as our mask to make fair comparison. We use a multiple pixel attack with backdoor strength (K=1). Additionally, we also compare with another strong attack \cite{Trojannn} with a different trigger generation and Trojan insertion technique than ours. We implement their Trojan generation technique on VGG-16 network. We did not use their data generation and denoising techniques as the assumption for our work is that the attacker has access to a set of random test batch. To make the comparison fair, we use similar trigger area, number of neurons and other parameters for all the baseline methods as well. 

\subsection{Evaluation Metrics}

\noindent\textbf{Test Accuracy (TA).} Percentage of test samples correctly classified by
the DNN model.

\noindent\textbf{Attack Success Rate (ASR).} Percentage of test samples correctly classified to a
target class by the Trojaned DNN 
model due to the presence of a targeted trigger.

\noindent\textbf{Number of Weights Changed ($w_b$):} The amount of weights which do not have exact same value between the model before attack(e.g, clean model) and the model after inserting the Trojan(e.g, attacked model).

\noindent\textbf{Stealthiness Ratio (SR)}
It is the ratio of (test accuracy $-$ attack failure rate) and $w_b$. 
\begin{equation}
    SR= \dfrac{TA-(100-ASR)}{w_b}
    = \dfrac{TA+ASR-100}{w_b}
\end{equation}
Now a higher SR indicates the attack does not change the normal operation of the model and less likely to be detected. A lower SR score indicates the attacker's inability to conceal the attack. 

\vspace{-1em}
\paragraph{Number of Bits Flipped ($n_b$)} The amount of bits attacker needs to flip to transform a clean model into an attacked model.
\vspace{-1em}
\paragraph{Trigger Area Percentage(TAP):} The percentage of area of the input image attacker needs to replace with trigger. If the size of the input image is $p \times q$ and the trigger size is $m \times m$ across each color channel then TAP can be calculated as:

\begin{equation}
TAP= \frac{m^2}{p \times q} \times 100 \%
\end{equation}

\section{Experimental Results}

\subsection{CIFAR-10 Results}
\cref{tab:classcifar} summarizes the test accuracy and attack success rate for different classes of CIFAR-10 dataset. Typically, an 8-bit quantized ResNet-18 test accuracy on CIFAR-10 is 91.9\%. We observe a certain drop in test accuracy for all the targeted classes. The highest test accuracy was 91.68\% when class 9 was chosen as the target class. 

Also, we find that attacking class 3,4 and 6 is the most difficult. Further. these target classes suffer from poor test accuracy after training. We believe that the location of the trigger may be critical to improving the ASR for class 3,4 and 6, since not all the classes have their important input feature at the same location. Thus, we further investigate different classes and trigger locations in the following discussion section. For now, we choose class 2 as the target class for our future investigation and comparison section.

\begin{table}[ht]
\centering
\caption{\textbf{CIFAR-10 Results}: vulnerability analysis of different class on ResNet-18. \textit{TC} indicates target class number. In this experiment we chose \textbf{$w_b$ to be 150} and trigger area was 9.76\% for all the cases.}
\label{tab:classcifar}
\begin{tabular}{cccccc}
\toprule
\textit{TC} & \textit{{\begin{tabular}[c]{@{}c@{}}TA\\ (\%)\end{tabular}}} & \textit{{\begin{tabular}[c]{@{}c@{}}ASR\\ (\%)\end{tabular}}} & \textit{TC} & \textit{{\begin{tabular}[c]{@{}l@{}}TA\\ (\%)\end{tabular}}} & \textit{{\begin{tabular}[c]{@{}l@{}}ASR\\ (\%)\end{tabular}}} \\ \midrule
0 & 91.05 & 99.20 & 5 & 89.93 & 95.91 \\
1 & 91.68 & 98.96 & 6 & 80.89 & 80.82 \\
2 & 89.38 & 93.41 & 7 & 86.65 & 85.40 \\
3 & 81.88 & 84.94 & 8 & 89.28 & 97.16 \\
4 & 84.35 & 89.55 & 9 & 91.48 & 96.40 \\
\bottomrule
\end{tabular}
\end{table}

By observing the Attack Success Rate (ASR) column, it would be evident that certain classes are more vulnerable to targeted bit Trojan attack than others. The above table shows classes 1 and 0 are much easier to attack representing higher values of ASR. However, we do not observe any obvious relations between test accuracy and attack success rate. But it is fair to say if the test accuracy is relatively high on a certain target class, it is highly probable that target class will result in a higher attack success rate as well.

\vspace{-0.5em}
\subsection{ImageNet Results:} 
We implement our Trojan attack on a large scale dataset such as ImageNet. For ImageNet dataset, we choose  $\textbf{TAP}$ of 11.2 \% and $w_b$ of 150. 

\begin{table}[ht]
\centering
\caption{ImageNet Results on ResNet-18 Architecture:}
\label{tab:imagenet}
\begin{tabular}{cccc}
\hline
\textit{\textbf{Method:}} & \textit{\textbf{TA}} & \textit{\textbf{ASR}} & \textit{\textbf{Wb}} \\ \hline
\textit{TBT} & \textit{69.14} & 99.98 & 150 \\
\end{tabular}
\vspace{-1em}
\end{table}
Our proposed TBT could achieve 99.98 \% attack success rate on ImageNet while maintaining clean data accuracy. Previous works \cite{gu2017badnets, Trojannn} did not report ImageNet accuracy in their works but by inspection, we claim our TBT requires modifying $\sim 3000 \times$ less number of parameters in comparison to Badnet \cite{gu2017badnets} which would require training of the whole network.

\subsection{Ablation Study.}
\paragraph{Effect of Trigger Area.}
In this section, we vary the trigger area (\textit{TAP}) and summarize the results in table \ref{tab:trs}. In this ablation study, we try to keep the number of weights modified from the clean model $w_b$ fairly constant (142$\sim$149). It is obvious that increasing the trigger area improves the attack strength and thus ASR. 

\begin{table}[t]
\centering
\caption{\textbf{Trigger Area Study:} Results on CIFAR-10 for various combination of targeted Trojan trigger area.}
\label{tab:trs}
\begin{tabular}{ccccc}
\toprule
\textit{{\begin{tabular}[c]{@{}c@{}}TAP\\ (\%)\end{tabular}}} & \textit{{\begin{tabular}[c]{@{}c@{}}TA\\ (\%)\end{tabular}}} & \textit{{\begin{tabular}[c]{@{}c@{}}ASR\\ (\%)\end{tabular}}} & $\textit{w}_b$ & $\textit{n}_b$ \\ \midrule

6.25 & 77.24 & 89.40 & 149 & 645 \\
7.91 & 86.99 & 92.03 & 143 & 626 \\
9.76 & 89.38 & 93.41 & 145 & 623 \\
11.82 & 90.56 & 95.97 & 142 & 627 \\
\bottomrule
\end{tabular}
\end{table}

One key observation is that even though we keep $w_b$ fairly constant, the values of $n_b$ changes based on the value of TAP. It implies that using a larger trigger area (e.g, TAP 11.82 \%) would require less number of vulnerable bits to inject bit Trojan than using a smaller TAP (e.g, 6.25 \%). Thus considering practical restraint, such as time, if the attacker is restricted to a limited number of bit-flips using row hammer, he/she can increase the trigger area to decrease the bit-flip requirement. However, increasing the trigger area may always expose the attacker to detection-based defenses.

\paragraph{Effect of $w_b$.}
Next, we keep the trigger area constant, but varying the number of weights modified $w_b$ in the table \ref{tab:nws}. Again, with increasing $w_b$, we expect $n_b$ to increase as well. Attack success rate also improves with increasing values of $w_b$.

\begin{table}[t]
\centering
\caption{\textbf{Number of weights study:} Results on CIFAR-10 for various combination of number of weights changed $w_b$ for ResNet-18.}
\label{tab:nws}
\begin{tabular}{ccccc}
\toprule
\textit{{\begin{tabular}[c]{@{}c@{}}TAP\\ (\%)\end{tabular}}} & \textit{{\begin{tabular}[c]{@{}c@{}}TA\\ (\%)\end{tabular}}} & \textit{{\begin{tabular}[c]{@{}c@{}}ASR\\ (\%)\end{tabular}}} & $\textit{w}_b$ & $\textit{n}_b$ \\ \midrule
9.76 & 79.54 & 79.70 & 10 & 37 \\
9.76 & 82.28 & 91.93 & 24 & 84 \\
9.76 & 81.80 & 89.45 & 48 & 173 \\
9.76 & 89.09 & 93.23 & 97 & 413 \\
9.76 & 89.38 & 93.41 & 145 & 623 \\
9.76 & 89.23 & 95.62 & 188 & 803 \\
\bottomrule
\end{tabular}
\vspace{-0.6em}
\end{table}

 We observe that modifying only 24 weights and 84 bits, TBT can achieve close to 91.93\% ASR even though the test accuracy is low (82.28\%). It seems that using a value of $w_b$ of around 97 is optimum for both test accuracy(89.09\%) and attack success rate(93.23\%). Increasing $w_b$ beyond this point is not desired for two specific reasons: first, the test accuracy does not improve much. Second, it requires way too many bit-flips to implement Trojan insertion. Our attack gives a wide range of attack strength choices to the attacker such as $w_b$ and $\textbf{TAP}$ to optimize between TA, ASR, and $n_b$ depending on practical constraints.

\subsection{Comparison to other competing methods.}
The summary of TBT performance with other baseline methods is presented in table \ref{tab:cmp}. For CIFAR-10 and SVHN results, we use the Trojan area of 11.82\% and 14.06 \%, respectively. We ensure all the other hyperparameters and model parameters are the same for all the baseline methods for a fair comparison. 

\begin{table}[t]
\caption{\textbf{Comparison to the baseline methods:} For both CIFAR-10 and SVHN we used VGG-16 architecture. Before attack means the Trojan is not inserted into DNN yet. It represents the clean model's test accuracy.}
\label{tab:cmp}
\scalebox{0.9}{
\begin{tabular}{cccccc}
\hline
\textit{{Method}} & \multicolumn{2}{c}{\textit{{\begin{tabular}[c]{@{}c@{}}TA\\ (\%)\end{tabular}}}} & \textit{{\begin{tabular}[c]{@{}c@{}}ASR\\ (\%)\end{tabular}}} & $\textit{w}_\textup{\textit{b}}$ & \multicolumn{1}{c}{\textit{{SR}}} \\ \cline{2-3}
\multicolumn{1}{c}{} & \multicolumn{1}{c}{\textit{{\begin{tabular}[c]{@{}l@{}}Before\\ Attack\end{tabular}}}} & \multicolumn{1}{c}{\textit{{\begin{tabular}[c]{@{}l@{}}After\\ Attack\end{tabular}}}} & \textit{{}} & \textit{\textbf{}} & \multicolumn{1}{l}{} \\ \midrule
\multicolumn{5}{c}{\textit{\textbf{CIFAR-10}}} & \multicolumn{1}{c}{} \\ \midrule
Proposed (TBT) & 91.42 & 86.34 & 93.15 & 150 & 0.56 \\
Trojan NN\cite{Trojannn}& 91.42 & 88.16 & 93.71 & 5120 & .015 \\
BadNet \cite{gu2017badnets} & 91.42 & 87.91 & 99.80 & 11M & 0 \\ \midrule
\multicolumn{5}{c}{\textit{\textbf{SVHN}}} & \multicolumn{1}{l}{} \\ \midrule
Proposed (TBT) & 99.56 & 73.87 & 73.81 & 150 & 0.32 \\
Trojan NN\cite{Trojannn} & 99.56 & 75.32 & 75.50 & 5120 & 0.009 \\
BadNet \cite{gu2017badnets} & 99.56 & 98.95 & 99.98 & 11M & 0 \\
\bottomrule
\end{tabular}
}
\vspace{-1.5em}
\centering
\end{table}

For CIFAR-10, the VGG-16 model before the attack has a test accuracy of 91.42\%. After the attack, for all the cases, we observe a test accuracy drop. Despite the accuracy drop, our method achieves a reasonable higher test accuracy of 86.34\%. Our proposed Trojan can successfully classify 93.15\% of test data to the target class. The performance of our attack is stronger in comparison to both the baseline methods. But the major contribution of our work is highlighted in $w_b$ column as our model requires significantly less amount of weights to be modified to insert Trojan. Such a low value of $w_b$
ensures our method can be implemented online in the deployed inference engine through row hammer based bit-flip attack. The method would require only a few bit-flips to poison a DNN. Additionally, since we only need to modify a very small portion of the DNN model, our method is less susceptible to attack detection schemes. Additionally, our method reports a much higher SR score than all the baseline methods as well. 

For SVHN, our observation follows the same pattern. Our attack achieves moderate test accuracy of 73.87 \%. TBT also performs on par with Trojan NN \cite{Trojannn} with almost similar ASR. As SVHN is a series of street numbers certain important locations of the features may vary based on target class and may contribute to the poor ASR as discussed in table \ref{tab:trigloc}. But BadNet \cite{gu2017badnets} outperforms the other methods with a higher TA and ASR on both CIFAR-10 and SVHN dataset. Again, The performance dominance of BadNet can be attributed to the fact that they assume the attacker is in the supply chain and can poison the training data. But practically, the attacker having access to the training data is a much stronger requirement. Further, it is already shown that BadNet is vulnerable to different Trojan detection schemes proposed in previous works \cite{wang2019neural,chen2018detecting}. Our proposed TBT requires $\sim 6M \times$ less number of parameter modification in comparison to BadNet.

\vspace{-1em}
\section{Discussion}

\begin{figure}[t]
    \centering
    \includegraphics[width=0.45\textwidth]{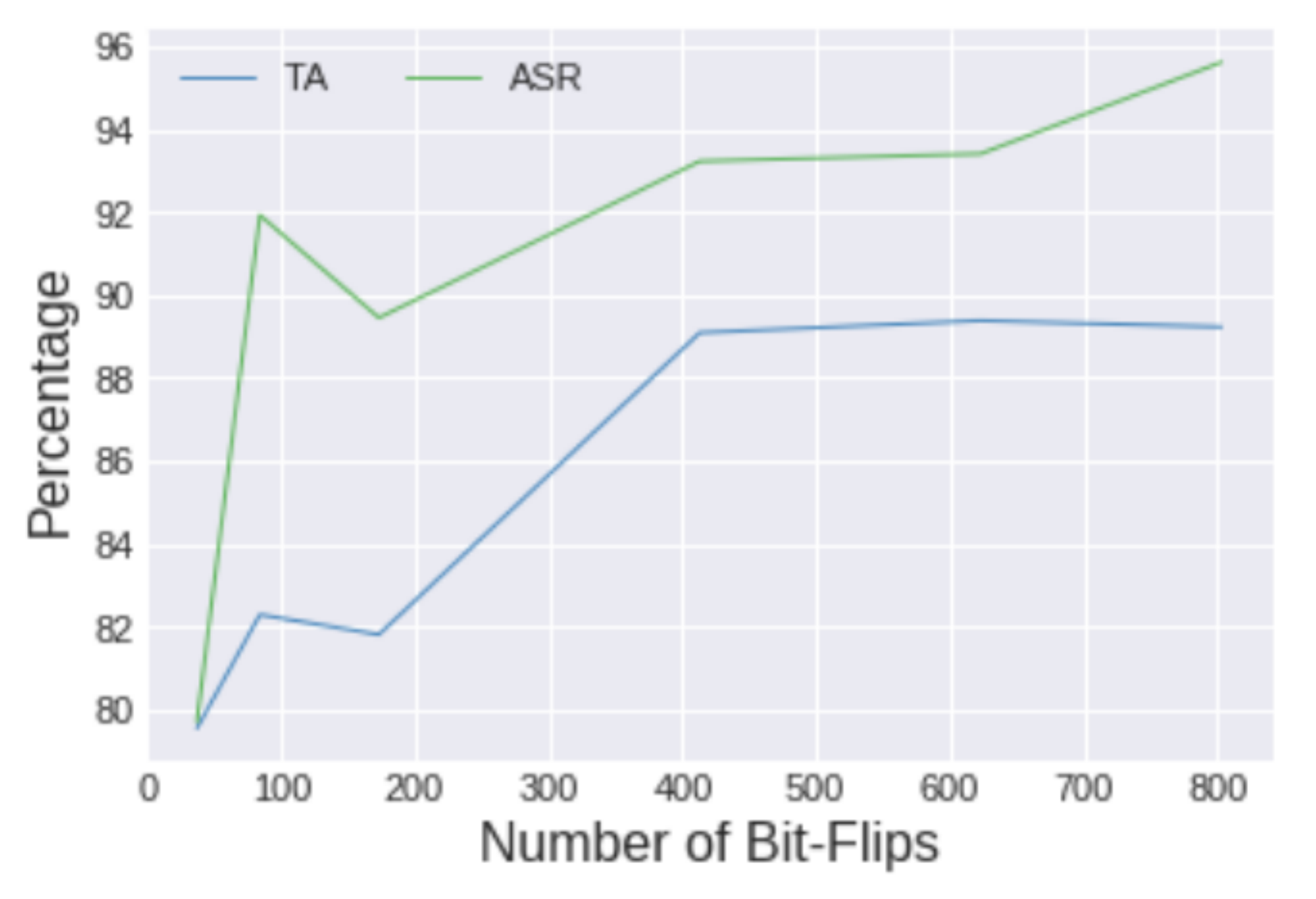}
    \caption{ ASR(Green) and TA(Blue)  vs number of bit flips plot. Only with 84 bit flips TBT can achieve 92 \% attack success rate.}
    \label{fig:bits}
    \vspace{-1em}
\end{figure}

\paragraph{Relationship between $n_b$ and ASR.}
We already discussed that an attacker, depending on different applications, may have various limitations. Considering an attack scenario where the attacker does not need to worry about test accuracy degradation or stealthiness, then he/she can choose an aggressive approach to attack DNN with a minimum number of bit-flips. Figure \ref{fig:bits} shows that just around 84 bit-flips would result in an aggressive attack. We call it aggressive because it achieves 92\% attack success rate (highest) with lower (82\%) test accuracy. Flipping more than 400 bits does not improve test accuracy, but to ensure a higher attack success rate.  

paragraph{Trojan Location and Target Class analysis:}
We attribute the low ASR of our attack in table \ref{tab:classcifar} for certain classes (i.e., 3,4,6,7) on trigger location. We conjecture that not all the classes have their important features located in the same location. Thus, keeping the trigger location constant for all the classes may hamper attack strength. As a result, for target classes 3,4,6 and 7 we varied the Trojan location to three places Bottom Right, Top Left and Center.

\begin{table}[t]
\centering
\caption{\textbf{Comparison of different trigger location:} We perform trigger position analysis on target classes 3,4,6,7 as we found attacking these classes are more difficult in table \ref{tab:classcifar}.$\textbf{TC}$ means target class.}
\label{tab:trigloc}
\begin{tabular}{|c|c|c|c|c|c|c|}
\hline
{TC} & \multicolumn{2}{c|}{\textit{{\begin{tabular}[c]{@{}c@{}}Bottom\\ Right\end{tabular}}}} & \multicolumn{2}{c|}{\textit{{\begin{tabular}[c]{@{}c@{}}Top\\ Left\end{tabular}}}} & \multicolumn{2}{c|}{\textit{{Center}}} \\
\hline
\multicolumn{1}{|l|}{} & \textit{{TA}} & \textit{{ASR}} & \textit{{TA}} & \textit{{ASR}} & \textit{{TA}} & \textit{{ASR}} \\ \hline
\textit{3} & \multicolumn{1}{c|}{81.88} & 84.94 & \multicolumn{1}{c|}{\textbf{90.40}} & \textbf{96.44} & 84.50 & 85.09 \\ \hline
4 & \multicolumn{1}{c|}{84.35} &   89.55 & \multicolumn{1}{c|}{86.52} & 95.45 & \textbf{89.77} & \textbf{98.27} \\ \hline
6 & 80.89 & 80.82 & \textbf{87.91} & \textbf{96.41} & 86.06 & 90.55 \\ \hline
7 & 86.65 &  85.40 & \textbf{86.80} & \textbf{91.91} & 83.33 & 86.88 \\ \hline
\end{tabular}
\end{table}

Table \ref{tab:trigloc} depicts that optimum trigger location for different classes is not the same. If the trigger is located at the top left section of the image, then we can successfully attack class 3,6 and 7. It might indicate that the important features of these classes are located near the top left region. For class 4, we found center trigger works the best. Thus, we conclude that one key decision for the attacker before the attack would be to decide the optimum location of the trigger. As the performance of the attack on a certain target class heavily links to the Trojan trigger location.
\vspace{-0.5em}

\begin{figure}[t]
    \centering
    \includegraphics[width=0.45\textwidth]{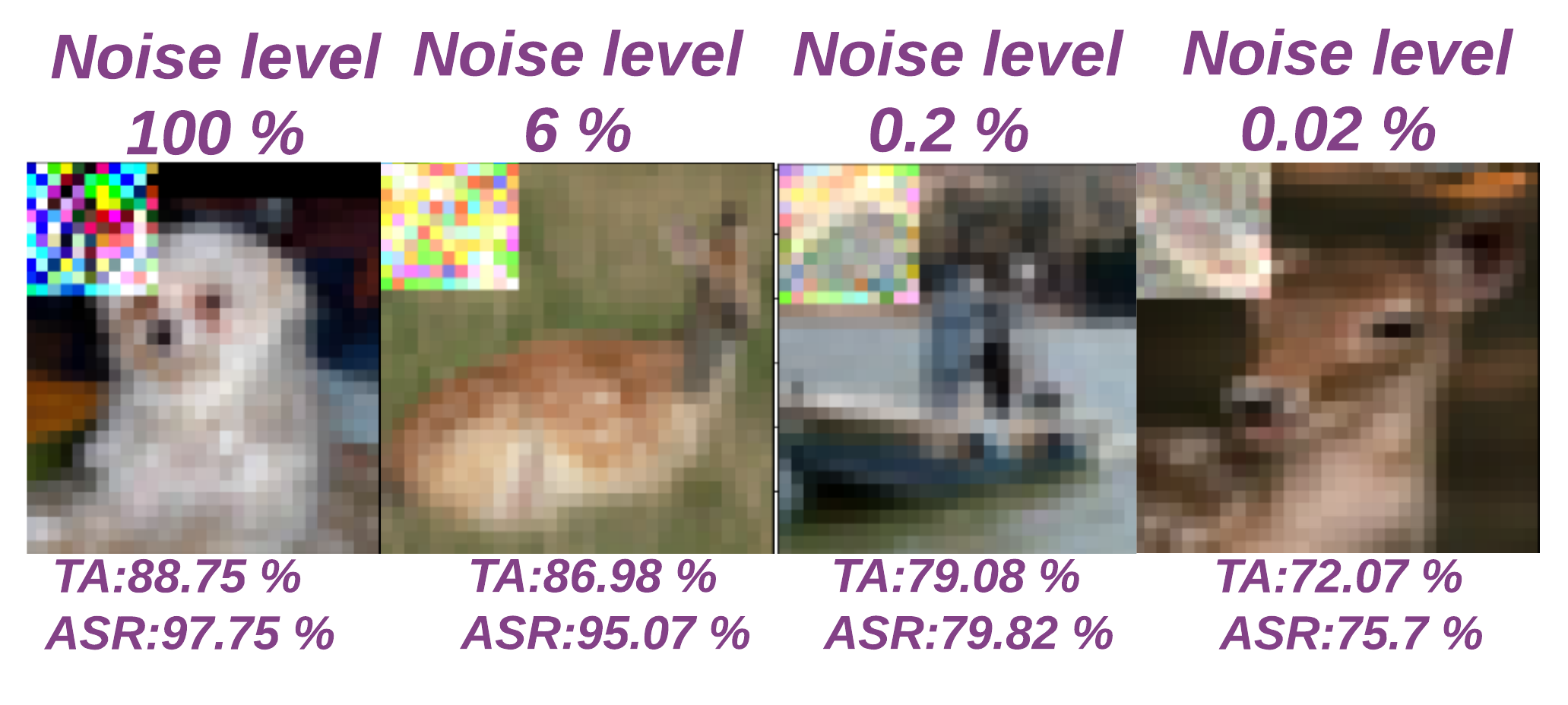}
    \caption{ Analysis of different noise level on CIFAR-10 dataset. TAP=9.76\%, $w_b$=150 and target class is 6. Noise Level: maximum amount of noise added to each pixel divided by the highest pixel value. We represent this number in percentage after multiplying by 100.}
    \label{fig:noise}
    \vspace{-1em}
\end{figure}
\vspace{-0.6em}
\paragraph{Trigger Noise level} In neural Trojan attack, it is common that the trigger is usually visible to human eye \cite{Trojannn,gu2017badnets}. Again, depending on attack scenario, the attacker may need to hide the trigger. Thus, we experiment to restrict the noise level of the trigger to 6\%, 0.2\% and .02\% in figure \ref{fig:noise}. Note that, the noise level is defined in the caption of figure \ref{fig:noise}. We find that the noise level in the trigger is strongly co-related to the attack success rate. The proposed TBT still fools the network with 79\% success rate even if we restrict the noise level to 0.2\% of the maximum pixel value. If the attacker chooses to make the trigger less vulnerable to Trojan detection schemes, then he/she needs to sacrifice attack strength.  

\vspace{-1em}
\paragraph{Potential Defense Methods}

\paragraph{Trojan detection and defense schemes}
\vspace{-1.5em}
As the development of neural Trojan attack accelerating, the corresponding defense techniques demand a thorough investigation as well. Recently few defenses have been proposed to detect the presence of a potential neural Trojan into DNN model \cite{Trojannn,chen2018detecting,liu2018fine,wang2019neural}. Neural Cleanse method \cite{wang2019neural} uses a combination of pruning, input filtering and unlearning to identify backdoor attacks on the model. Fine Pruning \cite{liu2018fine} is also a similar method that tries to fine prune the Trojaned model after the back door attack has been deployed. Activation clustering is also found to be effective to detect Trojan infected model \cite{chen2018detecting}. Additionally, \cite{Trojannn} also proposed to check the distribution of falsely classified test samples to detect potential anomaly in the model.
The proposed defenses have been successful in detecting several popular Trojan attacks \cite{Trojannn,gu2017badnets}. The effectiveness of the proposed defenses makes most of the previous attacks essentially impractical.  

However, one major limitation of these defenses is that they can only detect the Trojan once the Trojan is inserted during the training process/in the supply chain. None of these defenses can effectively defend during run time when the inference has already started. As a result, our online Trojan insertion attack makes TBT can be considered as practically immune to all the proposed defenses. For example, only the attacker decides when he/she will flip the bits. It requires significant resource overhead to perform fine-pruning or activation clustering continuously during run time. Thus our attack can be implemented after the model has passed through the security checks of Trojan detection. 

\paragraph{Data Integrity Check on the Model}
\vspace{-1.2em}
The proposed TBT relies on flipping the bits of model parameters stored in the main memory. One possible defense can be data integrity check on model parameters. Popular data error detection and correction technique to ensure data integrity are  Error-Correcting Code  (ECC) and Intel's SGX. However, row hammer attacks are becoming stronger to bypass various security checks such as ECC \cite{cojocar2019exploiting} and Intel's SGX \cite{gruss2018another}. Overall defense analysis makes our proposed TBT an extremely strong attack method which leaves modern DNN more vulnerable than ever. So our work encourages further investigation to defend neural networks from such online attack methods.

\paragraph{Our approach to Defend TBT}
\vspace{-1.2em}
In this work, we also investigate a different network architecture topology which may resist such a strong targeted attack better. An architecture with a complete different topology is Network In Network(NIN) \cite{lin2013network} which does not contain a fully-connected layer at the output and also utilizes global pooling. They propose global pooling as a regularizer which enforces feature maps to be confidence map of concepts. To conduct our attack with a last layer convolution layer like the Network in Network (NIN) architecture, we need to remove the last convolution layer and the average pooling layer to create the function g(x,$\theta$) described in the algorithm. The rest of the procedure will remain the same. We performed the experiment on NIN to confirm that our attack still achieves 99 \% success rate. But the clean test accuracy drops to 77 \% on CIFAR-10. We conjecture that the last layer average pooling may have further effect in weakening the attack stealthiness. Such a poor test accuracy may give defence techniques more chance to detect the presence of an attack. 
\vspace{-1.5em}
\section{Conclusion}
\vspace{-0.8em}
Our proposed Targeted Bit Trojan attack is the first work to implement neural Trojan into the DNN model by modifying small amount of weight parameters after the model is deployed for inference. The proposed algorithm enables Trojan insertion into a DNN model through only several bi-flips in computer main memory using row-hammer attack. Such a run time and online nerual Trojan attack puts DNN security under severe scrutiny. As a result, TBT emphasizes more vulnerability analysis of DNN during run time to ensure secure deployment of DNNs in practical applications.

\bibliographystyle{unsrt}
\bibliography{reference}

\end{document}